\documentclass[showpacs,twocolumn]{revtex4}
\usepackage{graphicx}
\begin{document}

\title{Different critical points of chiral and deconfinement phase transitions in (2+1)-dimensional fermion-gauge interacting model}
\author{Hong-tao Feng$^{1,4}$\footnote{Email:fenght@seu.edu.cn}, Feng-yao Hou$^{2,4}$, Yong-hui Xia$^3$, Jun-yi Wang$^{3}$ ,and Hong-shi Zong$^{3,4}$\footnote{Email:zonghs@chenwang.nju.edu.cn}}
\address{$^1$Department of Physics, Southeast University, Nanjing 211189, P. R. China}
\address{$^2$Institute of Theoretical Physics, CAS, Beijing 100190, P.R.
China}
\address{$^3$Department of Physics, Nanjing University, Nanjing 210093, P. R. China}
\address{$^4$State Key Laboratory of Theoretical Physics, Institute of Theoretical Physics, CAS, Beijing 100190, China}
\begin{abstract}

Based on the truncated Dyson-Schwinger equations for fermion and
massive boson propagators in QED$_3$, the fermion chiral condensate
and the mass singularities of the fermion propagator via the Schwinger function
are investigated. It is shown that the critical point of chiral
phase transition is apparently different from that of deconfinement
phase transition and in Nambu phase the fermion is confined only for small gauge boson mass.

\end{abstract}

\maketitle

 \maketitle
\section{introduction}
The chiral and deconfinement phase transitions of nonperaturbative
systems are important issues of continuous interests both
theoretically and experimentally. Although the mechanism is unknown,
the originally chiral symmetric system may undergo chiral phase
transition (CPT) into a phase with dynamical chiral symmetry
breaking (DCSB) which explains the origin of constituent-quark
masses in QCD and underlies the success of chiral effective field
theory \cite{a1,a2}. In the chiral limit, the order parameter of CPT
is defined via the fermion propagator
\begin{equation}\label{CDS}
\langle\bar\psi\psi\rangle=Tr[S(x\equiv0)]=\int\frac{\mathrm{d}^dp}{(2\pi)^d}\frac{4B(p^2)}{A^2(p^2)p^2+B^2(p^2)}.
\end{equation}
The two functions $A(p^2)$ and $B(p^2)$ in the above equation are
related to the inverse fermion propagator
\begin{equation}\label{S0}
 S^{-1}(p)= i\gamma\cdot pA(p^2)+B(p^2).
\end{equation}
The deconfinement phase transition is then related to the
observation of the free particle and also the corresponding
propagator. If the full fermion propagator has no mass singularity
in the timelike region, it can never be on mass shell and the free
particle can never be observed where the confinement
happens \cite{a3}. Accordingly, the appearance of the mass
singularity in the system directly implies deconfinement. So in this
way we can learn the deconfinement phase transition from the
analytic structure of the fermion propagator.

To indicate DCSB and confinement, it is very suggestive to study
some model that reveals the general nonperaturbative features while
being simpler. Three-dimensional quantum electrodynamics (QED$_3$)
is just such a model which has many features similar to quantum
chromodynamics (QCD), such as DCSB and confinement
\cite{a2,a3,a5,a6,a6a,a6b,a6c}. Moreover, its superrenormalization
obviate the ultraviolet divergence which is present in QED$_4$. Due
to these reasons, it can serve as a toy model of QCD. In parallel
with its relevance as a tool through which to develop insight into
aspects of QCD, QED$_3$ is also found to be equivalent to the
low-energy effective theories of strongly correlated electronic
systems. Recently, QED$_3$ has been widely studied in graphene
\cite{a7,a8,a9} and high-T$_c$ cuprate superconductors
\cite{a10,a11,a12,a13}.

The study of DCSB in QED$_3$ has been an active subject near 30
years since Appelquist \emph{et al}. found that DCSB vanishes when
the flavor of massless fermions reaches a critical number
$N_c\approx3.24$ \cite{a14}. They gain this conclusion by solving
the truncated Dyson-Schwinger equation (DSE) for the fermion
propagator in the chiral limit. Later, extensive analytical and
numerical investigations showed that the existence of DCSB in
QED$_3$ remains the same after including higher order corrections to
the DSE \cite{a15,a16}. On the other hand, the achievement in
research of the mass singularity and confinement in QED$_3$ is
caused by a paper of P. Maris who found that the fermion is confined
by the truncated DSE for the full fermion and boson propagators at
$N<N_c$ \cite{a3} where chiral symmetry is broken. This result might
imply that the existence of confinement and DCSB depend on the same
boundary conditions. Moreover, the authors of Ref. \cite{a2,a16a}
pointed out that restoration of chiral symmetry and deconfinement
are coincident owing to an abrupt change in the analytic properties
of the fermion propagator when a nonzero scalar self-energy becomes
insupportable.

Nevertheless, the above result will be altered when the gauge boson
acquires a finite mass $\zeta$ through the Higgs
mechanism \cite{a17,a18}. For a fixed
$N(<N_c)$ and with the increasing boson mass, the fermion chiral condensate falls and diminishes at a
critical value $\zeta_c$ (which, of course, depends on $N$) and then
chiral symmetry restores. Since DCSB and confinement are
nonperaturbative phenomena, both of them occur in the low energy region
and might disappear with the rise of boson mass. Therefore, it is very
interesting to investigate whether or not both phase transitions
occur at the same critical point in this case. In this paper, we
will adopt the truncated DSEs for the full propagators to study the
behaviors of the mass singularity and the fermion chiral condensate with
a range of gauge boson mass and try to answer this question.

\section{Schwinger function}

The Lagrangian for massless QED$_3$ in a general covariant gauge in
Euclidean space can be written as
\begin{equation}
\mathcal{L}=\bar{\psi}(\not\!{\partial}-i\mathrm{e}\not\!\!{A})\psi+\frac{1}{4}F^{2}_{\sigma\nu}+\frac{1}{2\xi}(\partial_{\sigma}A_{\sigma})^2,
\end{equation}
where the 4-component spinor  $\psi$ is the massless fermion field,
$\xi$ is the gauge parameter. This system has chiral symmetry and
the symmetry group is $U(2)$. The original $U(2)$ symmetry reduces
to $U(1)\times U(1)$  when the massless fermion acquires a nonzero
mass due to nonperaturbative effects. Just as mentioned in Sec. I,
the chiral symmetry is broken by the dynamical generation of the
fermion mass (here $N=1$). If one adopts the full boson propagator,
the results of Euclidean-time Schwinger function reveal that the
fermion propagator has a complex mass singularity and thus
corresponds to a nonphysical observable state \cite{a3} which means
the appearance of confinement. On the contrary, if the Schwinger
function exhibits a real mass singularity of the propagator, the
fermion is observable and the fermion is not confined
\cite{a19,a20}. Therefore, we also adopt this method to analyze
those nonperaturbative phenomena.

The Schwinger function can be written as
\begin{equation}\label{SF}
 \Omega(t)= \int\mathrm{d}^2\vec x\int\frac{\mathrm{d}^3p}{(2\pi)^3}e^{i(p_0t+\vec p\cdot \vec x)}\frac{M(p^2)}{p^2+M^2(p^2)}
\end{equation}
with $M(p^2)=B(p^2)/A(p^2)$. If there are two complex conjugate mass
singularities $m^*=a\pm ib$ associated with the fermion propagator,
the function will show
 an oscillating behavior
 \begin{equation}\label{SFC}
 \Omega(t)\sim e^{-at}\cos(bt+\phi)
 \end{equation}
for large (Euclidean) $t$. However, the system reveals a stable
observable asymptotic state with a mass $m$ for the fermion
propagator, then
\begin{equation}\label{SFDC}
    \Omega(t)\sim e^{-mt}\Rightarrow \lim
    _{t\rightarrow\infty}\ln\Omega(t)\sim -mt.
\end{equation}
By this way, the analysis of mass singularity can be used to
determine whether or not the fermion is confined. Since the Schwinger function
is determined by the fermion propagator and the DSEs provide us an
powerful tool to study it, we shall use the coupled gap equations to
calculate this function.

\section{truncated DSE}

Now let us turn to the calculation of $A(p^2)$ and $B(p^2)$.
These functions can be obtained by solving DSEs for the fermion
propagator,
\begin{equation}
S^{-1}(p)=S^{-1}_{0}(p)+\int\frac{\mathrm{d}^{3}k}{(2\pi)^{3}}[\gamma_{\sigma}S(k)\Gamma_{\nu}(p,k)D_{\sigma\nu}(q)],
\label{eq2}
\end{equation}
where $\Gamma_{\nu}(p,k)$ is the full fermion-photon vertex and
$q=p-k$. The coupling constant $\alpha=\mathrm{e}^2$ has dimension
one and provides us with a mass scale. For simplicity, in this
paper temperature, mass and momentum are all measured in unit of
$\alpha$, namely, we choose a kind of natural units in which
$\alpha= 1$. Form Eq. (\ref{S0}) and Eq. (\ref{eq2}), we obtain the
equation satisfied by $A(p^2)$ and $B(p^2)$
\begin{eqnarray}
A(p^{2})&=&1-\frac{1}{4p^2}\int\frac{\mathrm{d}^{3}k}{(2\pi)^{3}}Tr[i(\gamma p)\gamma_{\sigma}S(k)\Gamma_{\nu}(p,k)D_{\sigma\nu}(q)],\\
B(p^{2})&=&\frac{1}{4}\int\frac{\mathrm{d}^{3}k}{(2\pi)^{3}}Tr\left[\gamma_{\sigma}S(k)\Gamma_{\nu}(p,k)D_{\sigma\nu}(q)\right].
\end{eqnarray}
Another involved function $D_{\sigma\nu}(q)$ is the full gauge boson
propagator which is given by\cite{a17,a18}
\begin{equation}
D_{\sigma\nu}(q)=\frac{\delta_{\sigma\nu}-q_{\sigma}q_{\nu}/q^{2}}{q^{2}[1+\Pi(q^{2})]+\zeta^2}+\xi\frac{q_\sigma
q_\nu}{q^4},
\end{equation}
where $\Pi(q^{2})$ is the vacuum polarization for the gauge boson
which is satisfied by the polarization tensor
\begin{equation}
\Pi_{\sigma\nu}(q^{2})=-\int\frac{\mathrm{d}^{3}k}{(2\pi)^{3}}Tr\left[S(k)\gamma_{\sigma}S(q+k)\Gamma_{\nu}(p,k)\right]
\end{equation}
and $\zeta$ is the gauge boson mass which is acquired though Higgs
mechanism which happens when the gauge field interacts with a scalar
filed in the phase with spontaneous gauge symmetry breaking (Here,
we adopt the massive boson propagator to investigate the oscillation
behavior of Schwinger function in DCSB phase, more details about
Higgs mechanism in QED$_3$ can be found in Ref. \cite{a17,a18a}).

Using the relation between the vacuum polarization $\Pi(q^{2})$ and
$\Pi_{\sigma\nu}(q^{2})$,
\begin{equation}
\Pi_{\sigma\nu}(q^{2})=(q^2\delta_{\sigma\nu}-q_{\sigma}
q_{\nu})\Pi(q^{2}),
\end{equation}
we can obtain an equation for $\Pi(q^2)$ which has an ultraviolet
divergence. Fortunately, it is present only in the longitudinal part
and is proportional to $\delta_{\sigma\nu}$. This divergence can be
removed by the projection operator
\begin{equation}
\mathcal{P}_{\sigma\nu}=\delta_{\sigma\nu}-3\frac{q_{\sigma}q_{\nu}}{q^2},
\end{equation}
and then we obtain a finite vacuum polarization\cite{a5a,a6}.

Finally, we choose to work in the Landau gauge, since the Landau
gauge is the most convenient and commonly used one. Once the
fermion-boson vertex is known, we immediately obtain the truncated DSEs
for the fermion propagator and then analyze the deconfinement and
chiral phase transitions in this Higgs model.

\subsection{Rainbow approximation}

The simplest and most commonly used truncated scheme for the DSEs is
the rainbow approximation,
\begin{equation}
\Gamma_\nu\rightarrow\gamma_\nu,
\end{equation}
since it gives us rainbow diagrams in the fermion DSE and ladder
diagrams in the Bethe-salpeter equation for the fermion-antifermion
bound state amplitude. In the framework of this approximation, the
coupled equations for massless fermion and massive boson propagators
reduce to the three coupled equations for $A(p^2)$, $B(p^2)$ and
$\Pi(q^2)$,
\begin{eqnarray}\label{RE11}
A(p^2)&=&1+\int\frac{\mathrm{d}^{3}k}{(2\pi)^3}\frac{2A(k^2)(pq)(kq)/q^2}{p^2G(k^2)[q^2(1+\Pi(q^2))+\zeta^2]},\\
\label{RFp}
B(p^2)&=&\int\frac{\mathrm{d}^{3}k}{(2\pi)^3}\frac{2B(k^2)}{G(k^2)[q^2(1+\Pi(q^2))+\zeta^2]},\\
\label{Rpi1}
\Pi(q^2)&=&\int\frac{\mathrm{d}^{3}k}{(2\pi)^3}\frac{2A(k^2)A(p^2)}{q^2G(k^2)G(p^2)}\times\nonumber\\&&[2k^2-4(k\cdot
q)-6(k\cdot q)^2/q^2],
\end{eqnarray}
with $G(k^2)=A^2(k^2)k^2+B^2(k^2)$. By application of iterative
methods, we can obtain $A,~B$ and $\Pi$.
\subsection{Improved scheme for DSE}
To improve the truncated scheme for DSE, there are several attempts
to determine the functional form for the full fermion-gauge-boson
vertex \cite{a20a,a21,a22,a23}, but none of them completely resolve
the problem. However, the Ward-Takahashi identity
\begin{equation}
(p-k)_\nu\Gamma_{\nu}(p,k)=S^{-1}(p)-S^{-1}(k),
\end{equation}
provides us an effectual tool to obtain a reasonable ansatze for the
full vertex \cite{a24}.  The portion of the dressed vertex which is
free of kinematic singularities, i.e. BC vertex, can be written as,
\begin{eqnarray}
  \Gamma_\nu(p,k) &=& \frac{A(p^2)+A(k^2)}{2}\gamma_\nu +\frac{B(p^2)-B(k^2)}{p^2-k^2}(p+k)_\nu\nonumber\\
  &&+(\not\!p+\not\!k)\frac{A(p^2)-A(k^2)}{2(p^2-k^2)}(p+k)_\nu.
\end{eqnarray}
Since the numerical results obtained using the first part of the
vertex coincide very well with earlier investigations \cite{a16}, we
choose this one as a suitable ansatze
\begin{equation}\label{BCM1}
    \Gamma_\nu^{BC_1}(p,k)\simeq\frac{1}{2}\left[A(p^2)+A(k^2)\right]\gamma_\nu
\end{equation}
to be used in our calculation. Following the procedure in rainbow
approximation, we also obtain the three coupled equations for
$A(p^2),~B(p^2)$ and $\Pi(q^2)$ in the improved truncated scheme for
DSEs,
\begin{eqnarray}\label{E11}
&&A(p^2)=1+\int\frac{\mathrm{d}^{3}k}{(2\pi)^3}\frac{A(k^2)[A(p^2)+A(k^2)](pq)(kq)/q^2}{p^2G(k^2)[q^2(1+\Pi(q^2))+\zeta^2]},\\
\label{Fp}
&&B(p^2)=\int\frac{\mathrm{d}^{3}k}{(2\pi)^3}\frac{[A(p^2)+A(k^2)]B(k^2)}{G(k^2)[q^2(1+\Pi(q^2))+\zeta^2]},\\
\label{pi1}
&&\Pi(q^2)=\int\frac{\mathrm{d}^{3}k}{(2\pi)^3}\frac{A(k^2)A(p^2)[A(p^2)+A(k^2)]}{q^2G(k^2)G(p^2)}\times\nonumber
  \\&&~~~~~~~~~~~~~~~~~~~~[2k^2-4(k\cdot q)-6(k\cdot q)^2/q^2],
\end{eqnarray}

\section{numerical results}
\begin{figure}[htp!]
\includegraphics[width=0.235\textwidth]{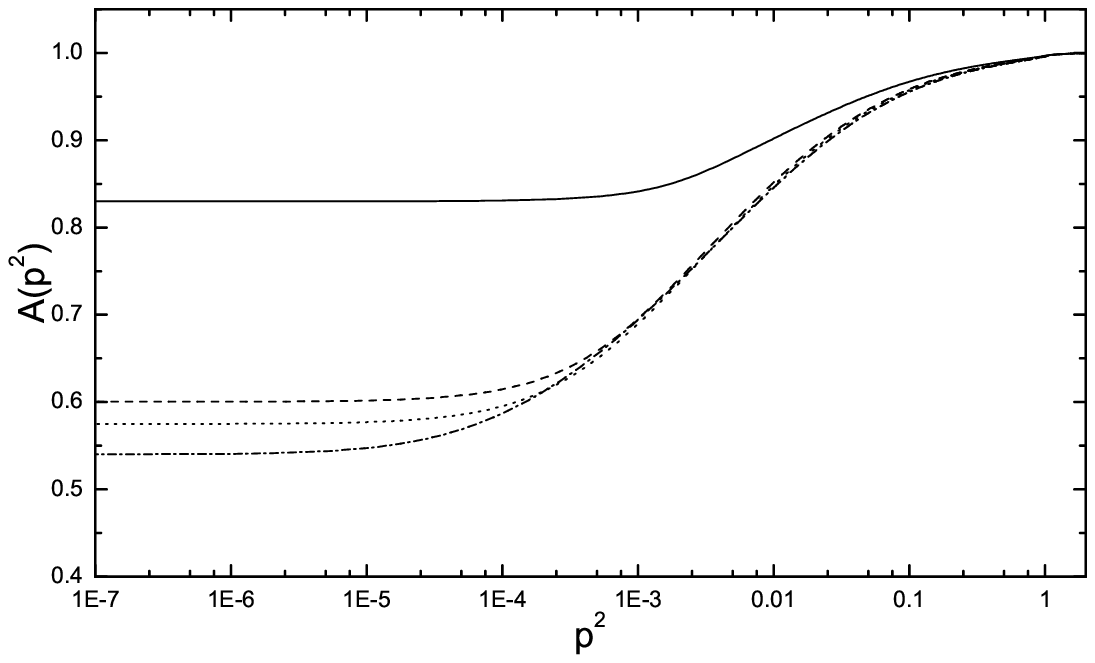}
\includegraphics[width=0.235\textwidth]{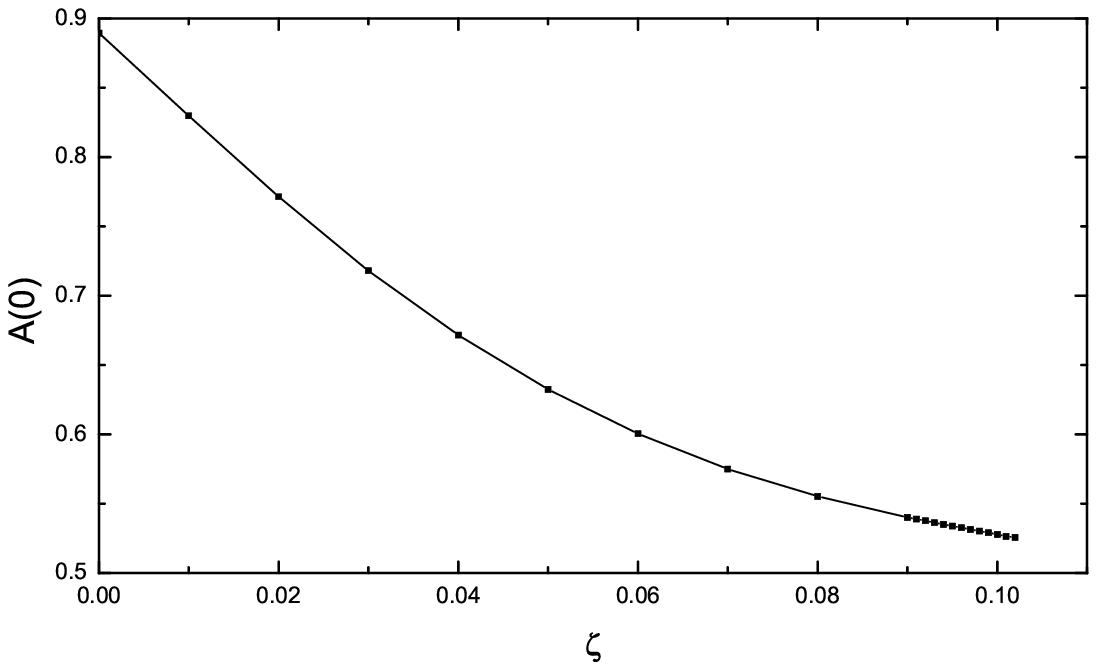}\\
\includegraphics[width=0.235\textwidth]{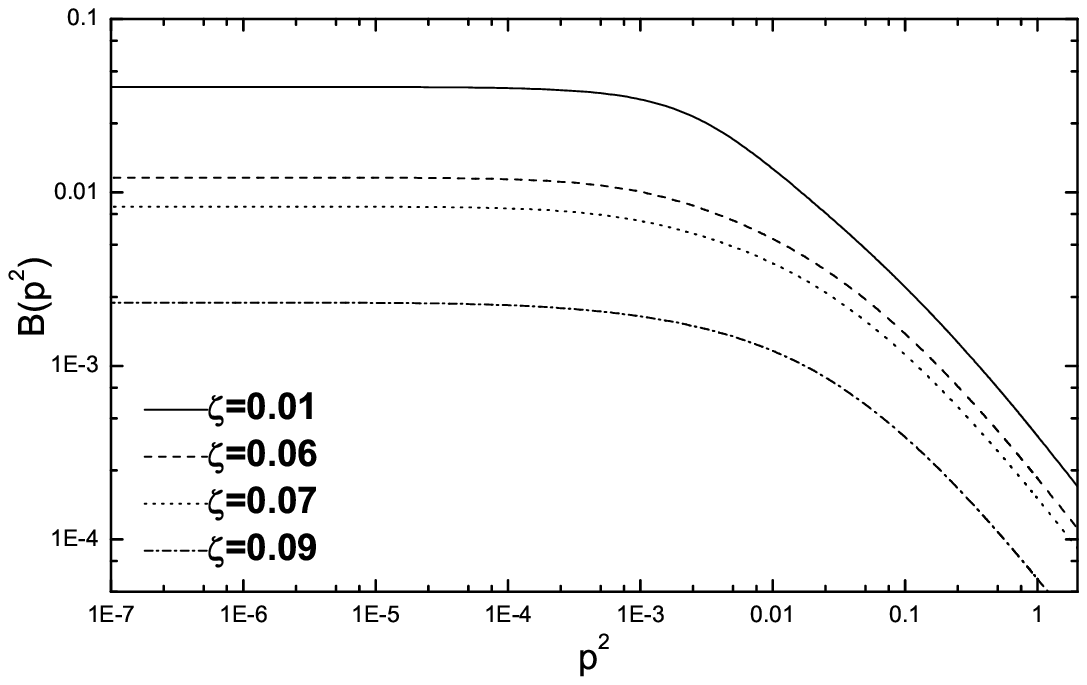}
\includegraphics[width=0.235\textwidth]{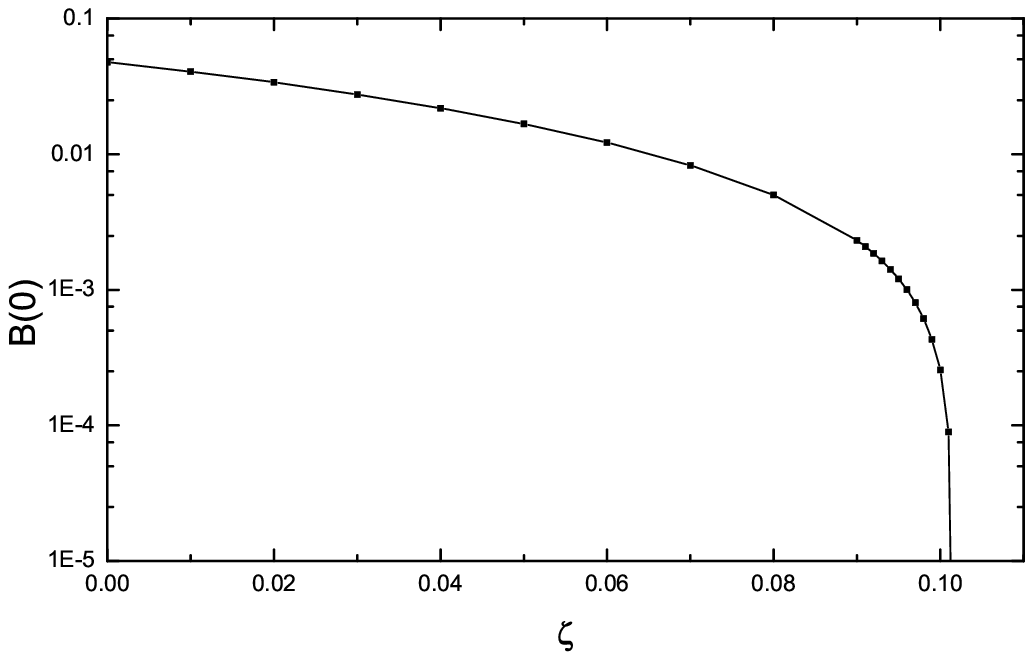}\\
\includegraphics[width=0.235\textwidth]{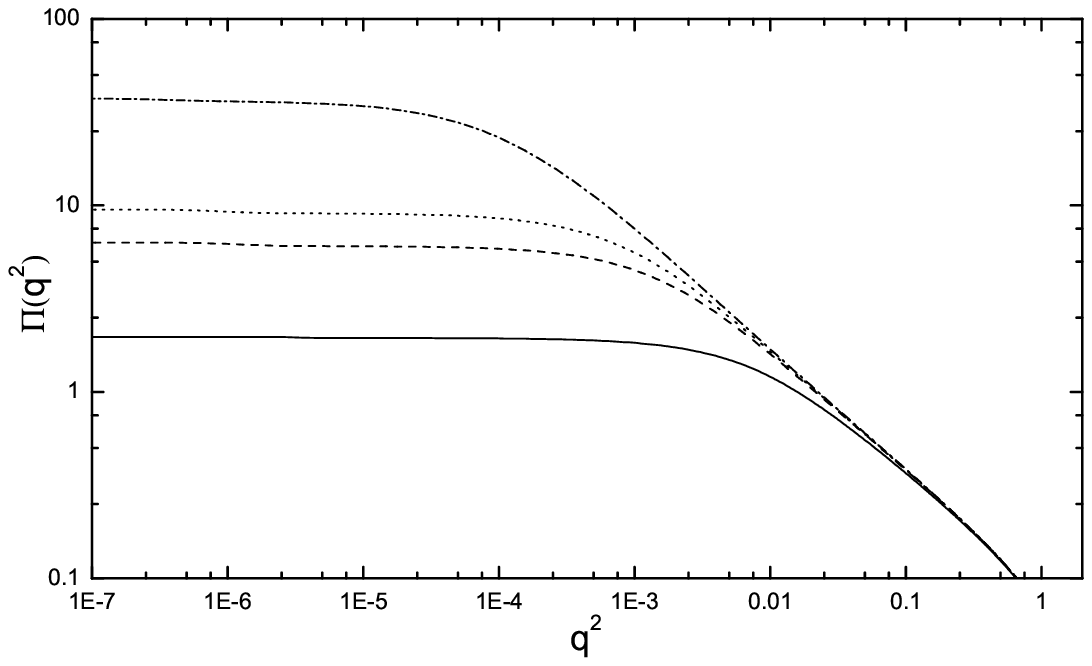}
\includegraphics[width=0.235\textwidth]{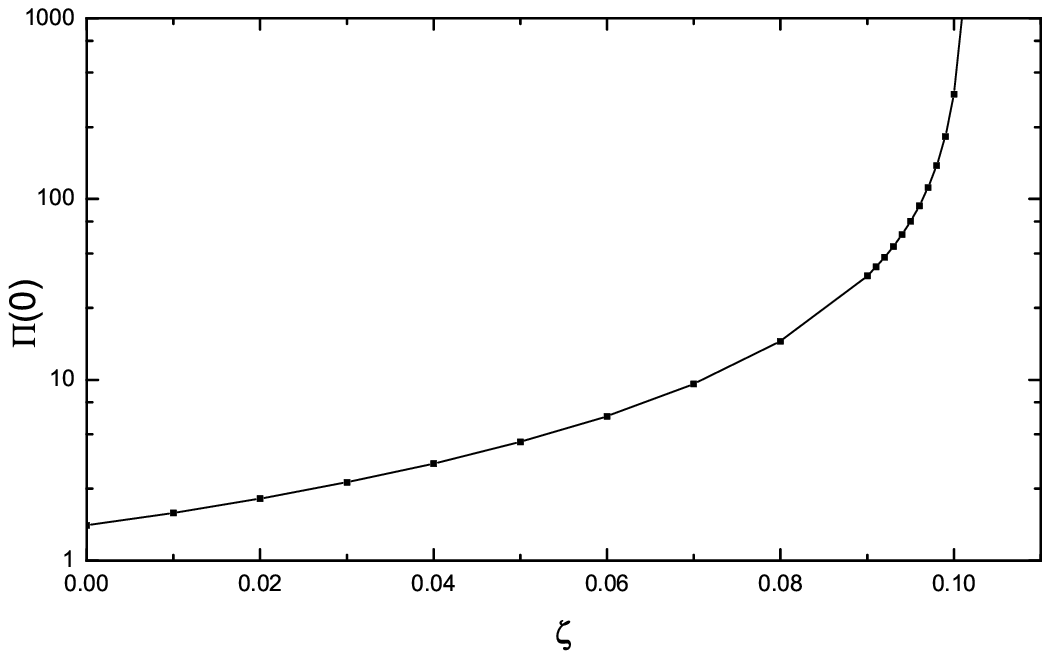}\\
\caption{The typical behaviors of $A(p^2),~B(p^2),~\Pi(q^2)$ (Left)
and their infrared values (Right) as functions of the boson mass in
DCSB phase.}\label{FIG1}
\end{figure}

After solving the above coupled DSEs  in rainbow approximation by
means of the iteration method, we can obtain the three function
$A,~B,~\Pi$ for the propagator and plot them in Fig. \ref{FIG1}.
From Fig. \ref{FIG1} it can be seen that $A(p^2)$ increases with
increasing momenta but almost equal to one at large $p^2$. In the
range of small momenta, it decreases but does not vanish when
$p^2\rightarrow0$. Both of the other two functions $B(p^2)$ and
$\Pi(q^2)$ decrease at large momenta but their rates of decreasing
are different. $B(p^2)$ decreases as rapidly as $\sim1/p^2$, while
$\Pi(q^2)$ decreases as rapidly as $\sim1/\sqrt{q^2}$. In addition,
all the three functions are constant in the infrared region. Thus,
we can obtain the values of the corresponding functions $A,~B$ and
$\Pi$ at zero momenta, which, as functions of the gauge boson mass
$\zeta$, are also shown in Fig. \ref{FIG1}. As $\zeta$ increases,
both $A(0)$ and $B(0)$ decrease, and $B(0)$ vanishes when $\zeta$
reaches a critical gauge boson mass $\zeta^R_c\approx0.102$, whereas
the function $\Pi(0)$ rises and diverges at the same critical boson
mass $\zeta^R_c$. Based on Eq. (\ref{CDS}), the critical boson mass
can be regarded as the point of chiral phase transition.

\begin{figure}[t]
\includegraphics[width=0.45\textwidth]{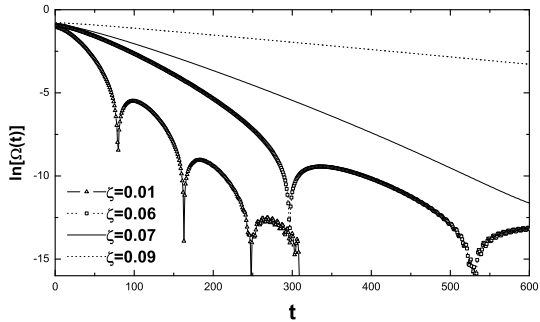}\\
\caption{Logarithm of the absolute value of the Schwinger function with
several $\zeta$ for the rainbow approximation.}\label{FIG2}
\end{figure}

Then, substituting the obtained $A$ and $B$ into Eq. (\ref{SF}), we
immediately obtain the behavior of the Schwinger function with
nonzero boson mass which is shown in Fig. \ref{FIG2}. At small
$\zeta$, the Schwinger function reveals its typical oscillating behavior
which illustrates the conjugate mass singularities like $m^*=a\pm ib$
\begin{eqnarray}
m^*&\sim&0.043\pm0.063i~\mathrm{at}~\zeta=0.01,\\
m^*&\sim &0.023\pm0.025i~\mathrm{at}~\zeta=0.06,
\end{eqnarray}
associated with the fermion propagator and thus the free particle
can never be observed where the fermion is confined. As the rise of
$\zeta$, the oscillating behavior remains but it vanishes at another
critical value $\zeta^R_{dc}\approx0.068$ and around which both of
the propagators do not exhibit any singularity.

Beyond $\zeta^R_{dc}$, the function  $\ln[\Omega(t)]\sim-mt$
 where  the stable asymptotic state of
the fermion is  observable
\begin{eqnarray}
 m&\approx&0.021~\mathrm{at}~\zeta=0.07,\\
 m&\approx&0.0041~\mathrm{at}~\zeta=0.09
\end{eqnarray}
and hence the deconfinement phase transition happens, but the DCSB
remains. With the enlargement of $\zeta$, the absolute slope of
$\ln[\Omega(t)]$ decreases and $m$ disappears at $\zeta^R_c$.

\begin{figure}[t]
  \includegraphics[width=0.45\textwidth]{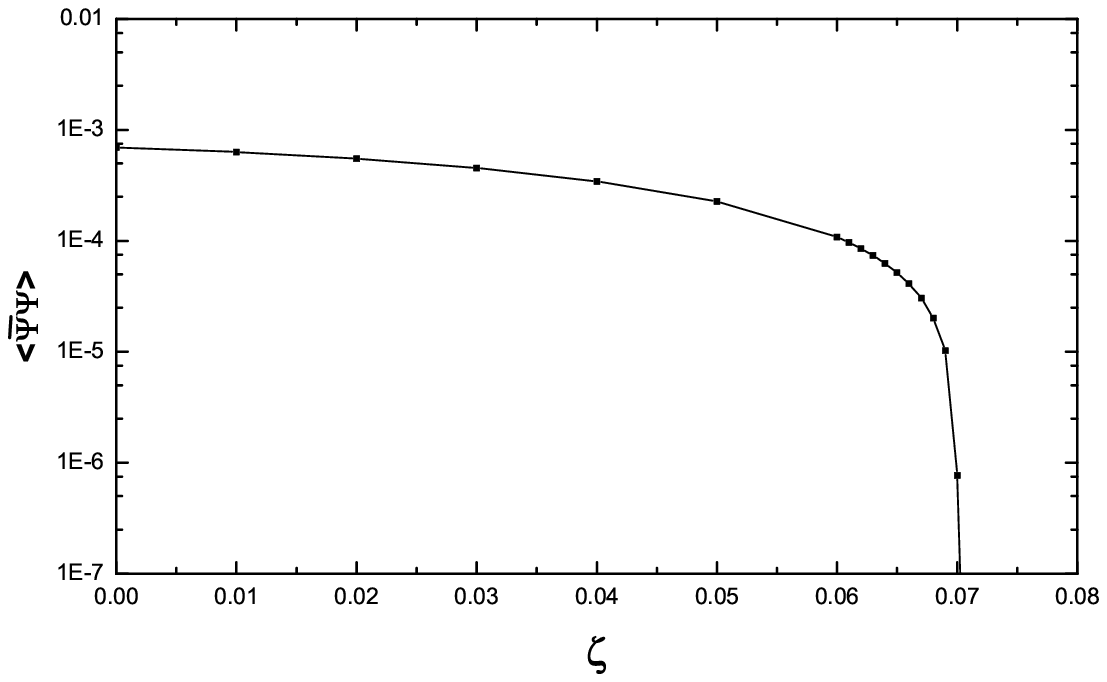}\\[-1cm]
\includegraphics[width=0.45\textwidth]{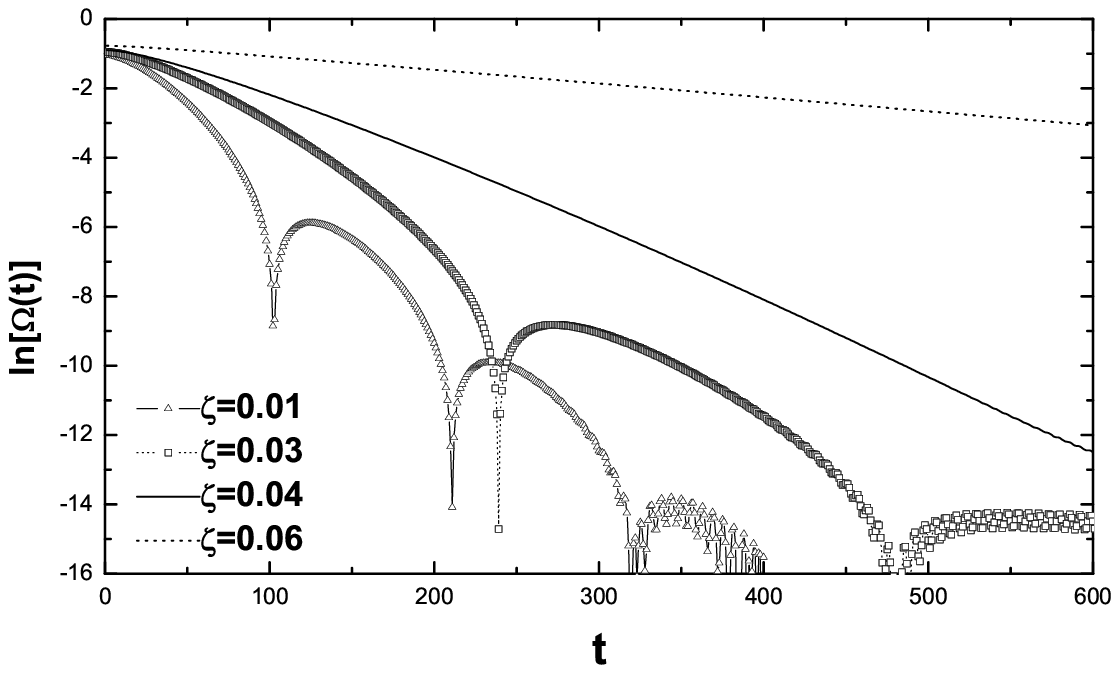}\\
\caption{The value of fermion chiral condensate (top) and the
logarithm (bottom) in the framework of BC$_1$ vertex with a range of
$\zeta$.}\label{FIG3}
\end{figure}
To validate the difference between $\zeta_c$ and $\zeta_{dc}$, we
also give the behavior of the Schwinger function beyond rainbow
approximation in Fig. \ref{FIG3}. In the BC$_1$ truncated scheme for
DSE, the oscillation of the Schwinger function only appears at small
$\zeta$, which denotes the existence of confinement, but it
disappears at $\zeta^{BC_1}_{dc}\approx0.038$, which exhibits that
deconfinement phase transition occurs but here
$\langle\bar\psi\psi\rangle\neq0$. As the rise of $\zeta$, the
Schwinger function shows the real mass singularity of the propagator
and chiral symmetry gets restored when the boson mass reaches
$\zeta^{BC_1}_{c}\approx0.071$.

\section{conclusions}
The primary goal of this paper is to investigate chiral and
deconfinement phase transition by application of an Abelian Higgs
model through a continuum study of the Schwinger function. Based on
the rainbow approximation of the truncated DSEs for the fermion
propagator and numerical model calculations, we study the behavior
of the Schwinger function and the fermion chiral condensate. It is
found that, with the rise of the gauge boson mass, the vanishing
point ($\zeta_{dc}$) of the oscillation behavior of the Schwinger
function is apparently less than that of the fermion chiral
condensate and each of the propagators \emph{does not} reveal any
singularity near $\zeta_{dc}$. To make know the difference between
the two critical points, we also work in an improved scheme for the
truncated DSEs and show that the above conclusion remains despite
the two critical numerical values alter. The result indicates that,
with the increasing gauge boson mass in the chiral model, the
occurrence of de-confinement phase transition is apparently earlier
than that of chiral phase transition.

\section{acknowledgements}

We would like to thank Prof. Wei-min Sun and Guo-zhu Liu for their
helpful discussions. This work was supported by the National Natural
Science Foundation of China (under Grant Nos. 11105029, 11275097 and
11205227) and the Fundamental Research Funds for the Central
Universities (under Grant No 2242014R30011).

\end{document}